\documentclass[11pt]{article}
\usepackage[dvips]{graphics, graphicx}
\usepackage{latexsym,epsf}
\usepackage{multicol}
\usepackage[centertags]{amsmath}
\fontencoding{OMS}\selectfont

\def\thebibliography#1{\setlength{\itemsep}{0pt}\setlength{\parsep}{0pt}
\refname{\small{\textbf{}}}
  \list
  {\arabic{enumi}.}{\settowidth\labelwidth{[#1]}\leftmargin\labelwidth
   \advance\leftmargin\labelsep
   \usecounter{enumi}}
\let\endthebibliography=\endlist}
\renewcommand\refname{\textbf{{REFERENCES}}}
\title{\textbf{Saturation of Coulomb sum rules in the $^6\rm{Li}$ case }}
\author{\textbf{\textit{A.Yu. Buki\footnote{\normalfont
Corresponding author. E-mail address: abuki@ukr.net.ua}, I.S.
Timchenko, N.G. Shevchenko  }}
\\
\emph{\small National Scientific Center "Kharkov Institute of
Physics and Technology", 61108, Kharkov, Ukraine}\\
{\small(Received June 8, 2011)}}
\begin{document}
\date{}
\maketitle
\begin{center}
\begin{minipage}{165 mm}
{\small
%----------------------   Abstract  --------------------------------%
The Coulomb sums $S_L(q)$ of the $^6$Li nucleus have been obtained
from electron scattering measurements at 3-momentum transfers $q =
1.125\: \div \:1.625$~fm$^{-1}$. It is found that at $q >
1.35$~fm$^{-1}$ the Coulomb sum of the nucleus becomes saturated:
$S_L(q) = 1$.
}\\
%-------------------------------------------------------------------%
PACS: 25.30.Fj, 27.20.+n \\
%-------------------------------------------------------------------%
\end{minipage}
\end{center}
\begin{multicols}{2}

\section{Introduction}
\label{intro}
The Coulomb sums (CS) are obtained from the treatment of data
 on electron scattering by atomic nuclei and can be used
 as an experimental data representation convenient for investigating
 some problems of the nuclear structure and the properties of
 intranuclear nucleons (\textit{e.g.}, see refs.~\cite{{1},{2},{3}}).
 The experimental CS were obtained for the most part at Saclay,
 Bates and SLAC Laboratories \cite{{2},{4},{5},{6}}.

  According to the sum rules, at sufficiently high momentum transfers
the CS must be a constant quantity \underline{equal} to $1$. The
experimental data show that with an increasing momentum transfer the
CS value of each of the nuclei studied also increases, and beginning
with  $q = 1.7 \div 2$~fm$^{-1}$  it becomes \underline{constant},
just as predicted by the sum rules. However, for the nuclei with the
atomic weight $A \geq 4$ at $q > 2$~fm$^{-1}$ the experimental CS
values are \underline{less} than $1$ (undersaturation of Coulomb sum
rules). To illustrate the CS behavior, fig.~1 shows the CS values
for the $^4\rm{He}$ nucleus.

  The problem of undersaturation of Coulomb sum rules has been
extensively discussed both in theoretical terms and in terms of
revising the measurement data and their processing. For example, the
revision of experimental CS at $q > 2$~fm$^{-1}$, made in paper~\cite{9}, has given the CS to be \underline{equal} to $1$ for the
$^{12}$C, $^{40}$Ca and $^{56}$Fe nuclei, while, on the contrary,
the revision of data in ref.~\cite{10} gave the CS to be
\underline{less} than $1$ for the $^{40, 48}$Ca, $^{56}$Fe,
$^{197}$Au, $^{208}$Pb and $^{238}$U nuclei. Theoretically, a
possible undersaturation of Coulomb sum rules may be assumed to be
due to modification of electromagnetic properties of the nucleon in
the nuclear matter environment.

  Apparently, the present-time experiment carried out at the Jefferson
Lab~\cite{11}, which is to verify the Saclay, Bates and SLAC
measurements on the $^{4}$He, $^{12}$C, $^{56}$Fe,  $^{208}$Pb, has
resulted from these longstanding discussions.

It is of importance to note that all the nuclei, for which the
experimental CS values were obtained, can be assigned to practically
unclustered nuclei, among which the nuclei having the atomic number
$A = 4\, \div \, 208$, can be classified with spherical nuclei.

The present paper is concerned with the CS of $^6$Li, \textit{i.e.},
the nucleus that is clustered and nonspherical.

\begin{center}
\vspace{-1cm}
\includegraphics[width=0.50\textwidth]{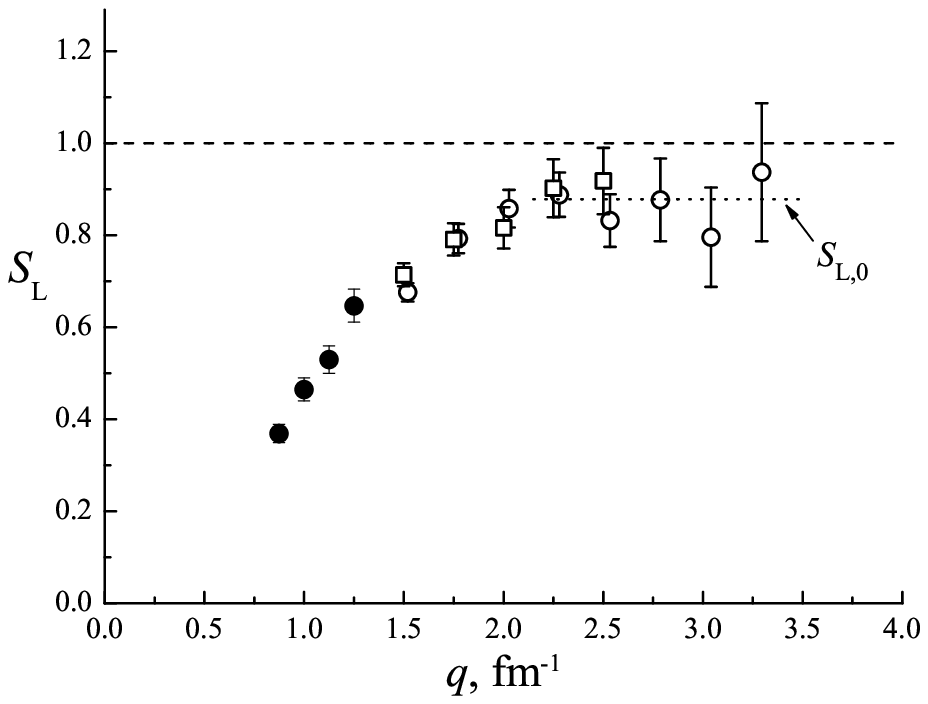}
\textbf{\it Fig.1.} {\it Coulomb sum of the $^4\rm{He}$ nucleus.
Open circles show the Saclay data \cite{4}, full circles - Kharkov
data \cite{7}, squares - Bates data \cite{8}, the dotted line
$S_{L,0}$ corresponds to the average $S_{L}(q)$ value in the range
of $q > 2$~fm$^{-1}$.}
\end{center}
%\vspace{-1cm}

\section{Terms and formulas}
\label{sec:1} One calls the Coulomb sum the zero moment of the
longitudinal response function (RF) of the atomic nucleus. The CS
can be represented as
\begin{equation}\label{1}
S_L(q)=\frac{1}{Z}\int_{\omega^+_{el}}^\infty\frac{R_L(q,\omega)}{G^2(q^2_\mu)\cdot\eta}\:
\rm{d}\omega. %\eqno(1)
\end{equation}
The subscript $\omega_{el}^+$ denotes the lower limit of integration
domain represented by the energy transfer that corresponds to
elastic electron scattering by the nucleus, though the form factor
of the process does not enter into the integral. The longitudinal
$R_L(q,\omega)$ and transverse $R_T(q,\omega)$ response functions
are related to the doubly differential cross section for inelastic
electron scattering from the nucleus $\frac{d^2\sigma}{d\Omega
d\omega}(\theta,E_0,\omega)$ by the known expression \cite{12},
which can be written as
$$
\frac{d^2\sigma}{d\Omega d\omega}(\theta,E_0,\omega)/\sigma_M(\theta,E_0) =
 $$
\begin{equation}\label{2}
  \frac{q^4_\mu}{q^4}R_L(q,\omega)+
\left[\frac{1}{2}\frac{q^2_\mu}{q^2}+\tan^2\frac{\theta}{2}\right]R_T(q,\omega).
%\eqno(2)
\end{equation}

In eqs.~(\ref{1}) and (\ref{2}) we use the following notation: $Z$ is the
nuclear charge; $G^2(q^2_\mu) = G_p^2(q^2_\mu) + (A-Z)\cdot
G_n^2(q^2_\mu)/Z$, $G_p(q^2_\mu)$ and $G_n(q^2_\mu)$ are,
respectively, the electrical proton and neutron form factors, which
were calculated by equations from ref.~\cite{13};  $\eta =
[1+q_\mu^2/(4M^2)]\times[1+q_\mu^2/(2M^2)]^{-1}$  is the correction
for the relativistic effect of nucleon motion in the nucleus, $M$ is
the proton mass; $\omega$,  $q$ and $q_\mu  = (q^2 -
\omega^2)^{1/2}$ are the energy, three- and four-momentum transfers
to the nucleus, respectively;  $E_0$ is the initial energy of
electron scattered through the angle  $\theta$;
$\sigma_M(\theta,E_0) = e^4 \cos^2(\theta/2) / [4 E_0^2
\sin^4(\theta/2)]$ is the Mott cross section; $e$ is the electron
charge. Here we use the effective 3-momentum transfer in the form $q
= \{ 4E_{eff}[E_{eff} - \omega] \sin^2(\theta/2) + \omega^2
\}^{1/2}$ , where \mbox{$E_{eff} = E_0 + 1.33 Ze^2
/\!<\!r^2\!>^{1/2}$} is the effective energy, in which the second
component takes into account the influence of the nuclear
electrostatic field on the incident electron \cite{14}, $<\!r^2\!>$
is the r.m.s. nuclear radius.

\section{Experiment and data processing}
\label{sec:2}
Spectra of electrons scattered by $^6$Li nuclei were
measured at the NSC KIPT electron linear accelerator LUE-300 at
initial energies $E_0$ = 130, 160, 177, 204, 233~MeV and at the
scattering angle $\theta = 160^{\circ}$, and also at \mbox{$E_0$ =
259 MeV} and $\theta = 53^{\circ}20',\, 60^{\circ}30',\,
61^{\circ}00',\, 68^{\circ}30',\, 77^{\circ}30',\, 94^{\circ}10'$.
The isotopic composition of targets in weight percent was 90.5\%
$^6$Li and 9.5\% $^7$Li. The content of other chemical elements in
the targets was less than 0.1\%. The target thickness along the
trajectory of electrons that hit the spectrometer was found to range
between 0.0024 and 0.0033 in radiation length units.

The scattered electrons are momentum analyzed by the spectrometer
SP-95 with a solid angle of $2.89 \times 10^{-3}$~sr and a
dispersion of 13.7~mm/percent. In the focal plane of the
spectrometer, the electrons are registered by eight scintillation
detectors, each having an energy acceptance of 1.4\%, and then
arrive at organic-glass Cherenkov radiators. The pulses from
photomultipliers of scintillation detectors and Cherenkov detectors
are registered by a coincidence circuit with a time resolution of 9~ns.

In spectral measurements, the background of accidental coincidences
of pulses from the scintillation/Cherenkov detectors was about or
less than 1\% of the effect value and was taken into account, while
the background measured in the absence of the target was one order
of magnitude lower. In the measured spectra, according to our
calculations and a few measurements with positrons, the background
contributed by $e^-e^+$-pairs from the target is insignificant if
present at all.

Before and after measuring each spectrum of electrons scattered by
$^6$Li, the peak of elastic electron scattering by $^{12}$C was
measured. Using the data of the measurements after their correction
for the radiation effects, the squared form factor F$_1^2(q)$ values
of the ground state of the $^{12}$C nucleus were found. These values
were used for normalizing our measured data for $^6$Li. Namely, the
data normalization factor was found as $k_{abs}$ =
F$_2^2(q)/$F$_1^2(q)$, where F$_2^2(q)$ stands for the squared form
factor of the $^{12}$C ground state measured in \cite{15} with a
systematic error of 0.4\%. Then, with the use of equations from
ref.~\cite{16}, the spectra of inelastic electron scattering by
$^6$Li were corrected for the radiation effects. Since the RF are
determined by inelastic electron scattering, the contribution from
the elastic scattering peak was subtracted from the $^6$Li$(e,e')$
spectra. Due to the fact that in our present measurements the energy
resolution in the neighborhood of the elastic scattering peak was
between 1.8 and 3.6~MeV, and the first excited-state energy of
$^6$Li was 2.18~MeV, then to subtract the elastic scattering
contribution from the spectrum of scattered electrons, we have used
the form factors of both the mentioned excited state and the ground
state measured in ref.~\cite{17}. The inelastic scattering cross
sections were divided by the corresponding Mott cross sections and
were averaged within 2~MeV intervals. In the group of spectra
measured at $\theta = 160^{\circ}$, the data were interpolated to
the $(q, \omega)$ values that corresponded to the spectra taken at
small scattering angles. At those $(q, \omega)$ values and with the
use of eq.~\ref{2}, the data were separated into $R_L(q, \omega)$
and $R_T(q, \omega)$ values. The obtained $R_{L}(q, \omega)$ values
were interpolated to the fixed 3-momentum transfer values: $q_c$ =
1.125, 1.250, 1.375, 1.500, 1.625~fm$^{-1}$. The interpolation
technique used here as well as some other additional details of the
measurements and data processing have been described in papers
\cite{{18},{19}}. Figure~2 shows the derived $R_L(q_c,\omega)$)
values.

\vspace{-0.5cm}
\begin{center}
\includegraphics[width=0.44\textwidth]{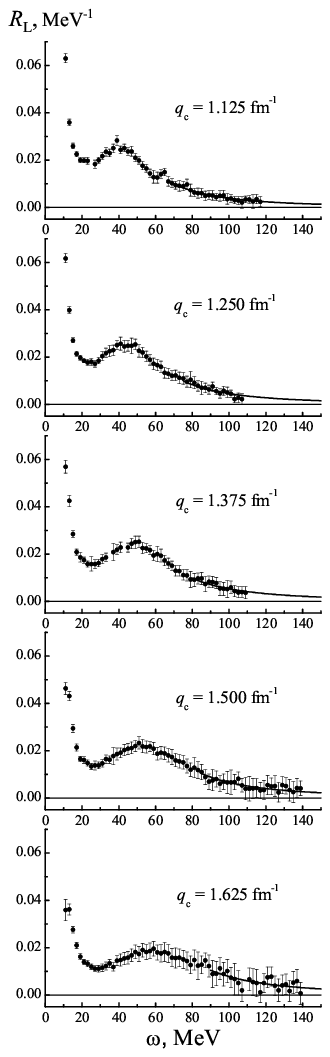}
\textbf{\it Fig.2.} {\it Longitudinal response function of the
$^6$Li nucleus. The curves represent the extrapolation of the
response functions (see the text).}
\end{center}

The attention is drawn to a relatively small (with regard to errors)
scatter of the experimental points. This is due to the fact that
each of the points is found through two interpolations of the
observed data, each interpolation smoothing the experimental data
sequences.
%It should be added that while at $\omega < 40$~MeV the
%interpolation was carried out along the lines lying in the plane of
%arguments $(q,\omega)$, at higher energy transfers the interpolation
%was performed over the surface formed by the experimental values of
%the function under consideration, and this resulted to smoother
%sequences of experimental data than in the first case.
 Note that the
data smoothing can also be observed for the experimental
$R_L(q_c,\omega)$ values obtained by other authors (\textit{e.g.}, see
refs.~\cite{{4},{8}}).

Relatively high $R_L(q_c,\omega)$ values in
the region of low $\omega$ are explained by the contribution from
the excitation of low-lying nuclear states, whose peaks have merged
because of a low energy resolution of measurements.

As regards the analysis of experimental RF, it would be of great
interest to compare them with current theoretical calculations.
However, by now, the modern calculations of RF have been made only
for the nuclei with $A \leq 4$ (\textit{e.g.}, see ref.~\cite{20}),
and for heavier nuclei these calculations are only projected.

To calculate the CS (eq.~\ref{1}), it is necessary to extrapolate the RF
to the high-energy transfer region. For this purpose, it is common
practice to use an exponential or a power function (\textit{e.g.}, see
refs.~\cite{6} and \cite{4}, respectively). The exponential
function, like the power function in some cases, as applied to the
RF, is considered as empirical. However, the RF extrapolation with
the power function has been substantiated in theoretical papers
\cite{{2},{21}} and \cite{22}, and, as applied to the experimental
RF values, it has been analyzed in ref.~\cite{23}. According to
refs.~\cite{{2},{21}}, the extrapolation power function has the form
\begin{equation}\label{3}
R^{\alpha}(q,\omega \rightarrow \infty) = C_{\alpha}(q)\cdot \omega'^{-\alpha},
\end{equation}
where $C_{\alpha}(q)$ is the fitting parameter, $\alpha$ is either
the fitting or the calculated parameter, $\omega'= \omega -
q^2/(2AM)$ is the c.m.s. energy transfer.

According to the calculations of ref.~\cite{21}, the parameter
$\alpha$ is equal to 2.5 and is independent of the momentum
transfer. With the use of the free parameters $C_{\alpha}(q)$  and
$\alpha$, from the fit of the function $R^{\alpha}(q,\omega
\rightarrow \infty)$ to the experimental RF of $^6$Li, obtained in
the high $\omega$ region, we have found $\alpha = 2.56\, \pm \,
0.06$. Then, using this $\alpha$ value for each $R_L(q_c,\omega)$ we
calculated  $C_{\alpha}(q)$. Using eq.~\ref{1}, the experimental
$R_L(q_c,\omega)$ and the function $R^{\alpha}(q,\omega \rightarrow
\infty)$ with the parameters found, we have obtained the Coulomb sum
$S_{L}(q)$ values. Besides, the Coulomb sums $S_L'(q)$ were obtained
through the use of exponential extrapolation of the type
\begin{equation}\label{4}
R^{\beta}(q,\omega \rightarrow \infty) = C_{\beta}(q)\cdot e^{-\beta(q) \cdot \omega'},
\end{equation}
with the fitting parameters $C_{\beta}(q)$ and
$\beta(q)$. Here, unlike the RF extrapolation with the power
function, the values of the two parameters $C_{\beta}$ and
$\beta$ are dependent on the momentum transfer.

The $S_L(q)$ and $S_{L}'(q$) values found here are presented in
table~\ref{tab:1} and fig.~3.
%\vspace{-0.5cm}
\begin{center}
\includegraphics[width=0.45\textwidth]{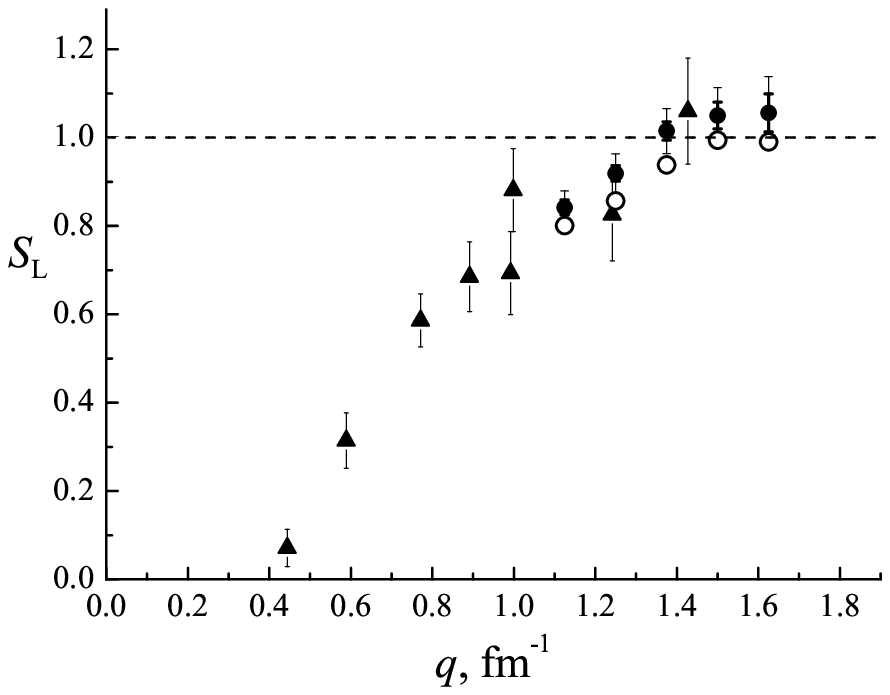}
\textbf{\it Fig.3.} {\it Coulomb sum of the $^6$Li nucleus.
Triangles show the CS values from ref.~\cite{25}; full circles and
open circles show the present data obtained with the use of power
and exponential extrapolations, respectively. The data denoted by
full circles include minor errors (statistical only) and major
errors (statistical plus systematic).}
\end{center}

\section{Significance of some corrections and
the errors for the CS} \label{sec:3}

In our present measurements the targets comprise 9.5\% $^7$Li by
weight. Earlier, we have made preliminary processing of measurements
on the targets consisting of 93.8\% $^7$Li and 6.2\% $^6$Li by
weight \cite{24,Li7arxiv}.

\begin{center}
\includegraphics[width=0.5\textwidth]{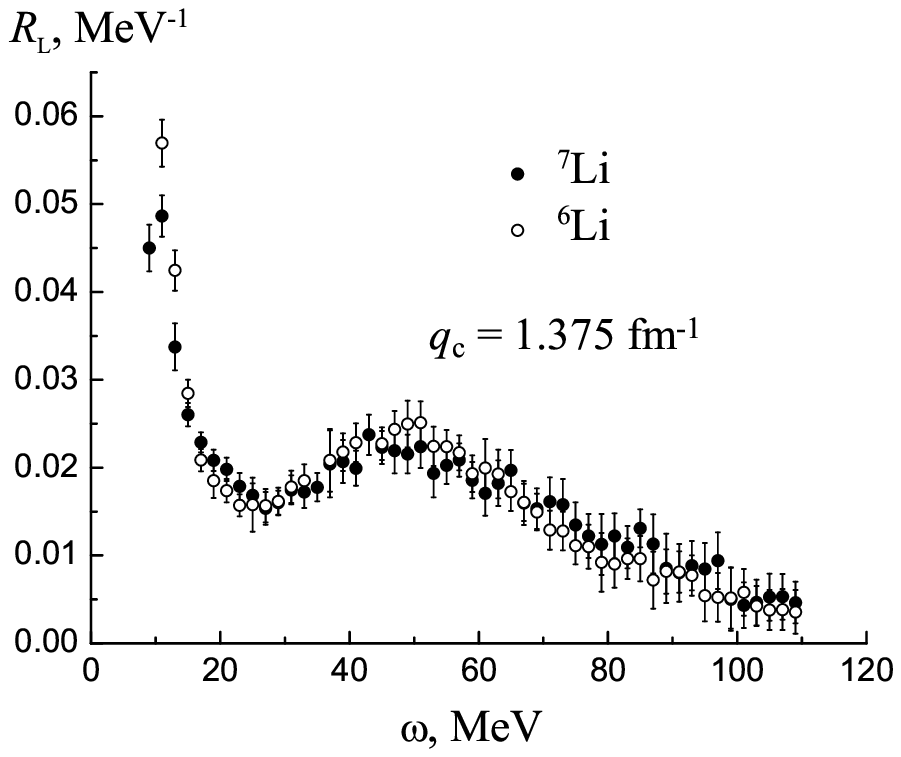}
\textbf{\it Fig.4.} {\it Longitudinal response functions of lithium
isotopes.  Full circles show the data of the present work for
$^6$Li; open circles - the data for $^7$Li taken from
ref.~\cite{24}.}
\end{center}

The RF of $^7$Li found in ref.~\cite{24,Li7arxiv} are close to the
RF of $^6$Li found here (see fig.~4),  and the $S_L(q)$ valuesof
$^7$Li are not different (to an accuracy of experimental errors)
from $S_L(q)$ values of $^6$Li. This implies that the presence of
$^7$Li impurity in the $^6$Li target exerts no essential effect on
the obtained $S_L(q)$ values of $^6$Li.

\textit{The correction contributions} to the $S_L(q)$ values
obtained are as follows:
\begin{description}
    \item [i)] the use of calculation of the electrical form factor
    of the proton, $G_p(q_{\mu}^2)$, by equations of ref.~\cite{13}
     instead of the traditional dipole formula, gives $(1.2 \div
     2.4)\%$;
    \item [ii)] $\eta$ in eq.~\ref{1} makes $(1.4 \div 2.8)\%$;
    \item [iii)] taking into account the nuclear Coulomb field
    effect on the momentum transfer of the incident electron is about
    1\%;
    \item [iv)] calculation of the electrical form factor of the neutron,
    $G_n(q_{\mu}^2)$ gives about 0.1\%.
\end{description}

\textit{The contributions to the systematic error} of $S_L(q)$
values, which come from the errors in:
\begin{description}
    \item [a)] normalization of data, including the measurement errors
    of the ground-state form factor of the $^{12}$C nucleus in the present
    work and in ref.~\cite{15} - $(1.2 \div 2.5)\%$;
    \item [b)] procedure of radiation correction of spectra - 2\%;
    \item [c)]  determination of thickness of $^6$Li targets employed
    in the measurements at $\theta = 160^{\circ}$ and at
    $\theta \leq 94^{\circ}10'$ - $(0.6 \div 1.3)\%$;
    \item [d)] interpolation procedures - up to 1\%;
    \item [e)] procedure of determination of the parameter $\alpha$ from the fit of eq.~\ref{3} to the experimental RF - $(0.6 \div 0.9)\%$;  %calculation of the parameter $\alpha$ in     Eq.~(3) - $(0.6 \div 0.9)\%$;
\end{description}

\textit{The contributions to the statistical error} of $S_L(q)$
values, which come from statistical errors of:
\begin{description}
    \item [f)] experimental $R_L(q_c,\omega)$ - $(1.7 \div 3.3)\%$;
    \item [g)] derivation of the parameter $C_{\alpha}(q)$ in
    eq.~\ref{3}  - $(1.2 \div 2.5)\%$.
\end{description}

To determine the contribution from correction iii) and the errors
(a, b, c, d, f), eq.~\ref{2} was used. The values of systematic ($\Delta
S_{L,syst}$) and statistical ($\Delta S_{L,stat}$) errors, given in
table~\ref{tab:1}, are the quadratic sums of the above-mentioned
contributions.

% For tables use
\begin{table*}
\caption{Coulomb sums $S_L$ and $S'_L$ of the $^6$Li nucleus
determined from the measurements up to the energy transfer
$\omega_{max}$ with the use of the power and the exponential
functions, respectively, for RF extrapolation. The
$S_{L,\,tail}/S_L$ and $S_{L,\,tail}'/S'_L$ ratios represent the
extrapolation fraction in the $S_L$ and $S'_L$ values. The errors of
$S_L$ and $S'_L$ are practically the same. }
\label{tab:1}       % Give a unique label
% For LaTeX tables use
\begin{center}
\begin{tabular}{|c|c|c|c|c|c|c|c|} \hline
$q$, fm$^{-1}$ & $\omega_{max}$, MeV & $S_L$& $\Delta S_{L,\, stat}$
& $\Delta S_{L,\, syst}$ & $S_{L,\, tail}/S_L$ & $S_L'$ & $S_{L,\,
tail}'/S'_L$\\\hline
 1.125 & 118 & 0.842 & 0.017 & 0.022 & 0.088 & 0.801 & 0.040 \\\hline
 1.250 & 108 & 0.919 & 0.018 & 0.026 & 0.117 & 0.857 & 0.052 \\\hline
 1.375 & 110 & 1.015 & 0.021 & 0.030 & 0.127 & 0.938 & 0.056 \\\hline
 1.500 & 140 & 1.050 & 0.030 & 0.033 & 0.105 & 0.994 & 0.056 \\\hline
 1.625 & 140 & 1.056 & 0.043 & 0.039 & 0.130 & 0.990 & 0.075 \\\hline
  \end{tabular}
  \end{center}
\end{table*}

\section{Discussion and conclusions}
\label{sec:4} The CS values obtained in the present study can be
supplemented with the data from ref.~\cite{25}. The CS of $^6$Li has
been denoted there as $\sigma_l(q)$, and is related to the
present-day definition of the Coulomb sum by the expression $S_L(q)
= \sigma_l(q)/G^2_p(q^2)$. The CS values of ref.~\cite{25}
transformed in this way are shown in fig.~3. It can be seen from the
figure that the function $S_L(q)$ for $^6$Li is different from
$S_L(q)$ for other nuclei (\textit{e.g.}, see $S_L(q)$ for $^4$He in
fig.~1).

Let us consider some special features of the CS for $^6$Li.

{\bf{5A.}} The data of Saclay and Bates Laboratories show an
increase in the CS up to $q = 1.7 \div 2$~fm$^{-1}$ for the nuclei
studied with $A = 4 \div 56$. As it can be seen from fig.~3, at $q
\leq 1.4$~fm$^{-1}$ the function $S_L(q)$ for $^6$Li attains the
range of constant values.\footnote{\normalfont Note that the special
feature of the CS for $^6$Li discussed here could also be seen in
the data of ref.~\cite{25}. However, in 1977, when that work was
published, there was no systematics of the CS data for a number of
nuclei (the data appeared only in the eighties) and it was
impossible to make any reasonable comparison between the CS of
different nuclei.} Relying on papers~\cite{{25},{26}}, it can be
demonstrated that if $S_L(q)$ of $^6$Li took on the constant value
at higher momentum transfers (as in the case of other nuclei), then
the clusterization in this nucleus would be small or absent.
However, that is not the case.

{\bf 5B.}
 For momentum transfers, at which the $S_L(q)$ values are
constant to an accuracy of experimental errors, we denote the
average CS as $S_{L,0}$. In the range of $q = 1.375 \div
1.625$~fm$^{-1}$ for $^6$Li we have $S_{L,0} = 1.031 \pm 0.016 \pm
0.034$, where the given errors are statistical and systematic,
respectively.

The found result shows the CS saturation, that corresponds to the
viewpoint of paper~\cite{9}. It has been stated there that the CS
undersaturation of the nuclei with $A \geq 4$, observed in the
Saclay and Bates experiments, was the result of error in the data
analysis.

If to take for granted the phenomenon of CS undersaturation
($S_{L,0} < 1$), revealed in the previously investigated nuclei with
$A \geq 4$, then $S_{L,0} = 1$ for $^6$Li falls out of the
systematics of the effect.

Here it should be noted that if the exhaustion or underexhaustion of
Coulomb sum rules is dealt with, it is generally assumed that the
$S_L(q)$ value is virtually fully determined by the cross section
for quasielastic electron scattering (QES) from intranuclear
nucleons. This takes place at $q \geq 2 $~fm$^{-1}$. The $S_{L,0}$
plateau of $A < 208$ nuclei (except $^6$Li) is observed at $q$
ranging from $1.7 \div 2$~fm$^{-1}$ to $3.5$~fm$^{-1}$. In the
$^6$Li case, this plateau begins at $q = 1.4$~fm$^{-1}$, and in the
measured range of momentum transfers ($q = 1.4 \div  1.6$~fm$^{-1}$)
the contribution of QES to $S_L(q)$ makes about 90\%. It is believed
that after reaching the plateau the $S_L(q)$ value of $^6$Li remains
constant in the region of high momentum transfers, too, as it is
observed in the case of other previously investigated nuclei with $A
< 208$.

 It appears of interest to consider this case ($^6$Li $S_L(q)
= 1$ at $q > 1.6$~fm$^{-1}$) from the standpoint of the hypothesis
about undersaturation of the Coulomb sum rules.

The undersaturation of Coulomb sum rules can be explained by the
modification of intranuclear nucleons. A prerequisite to the nucleon
modification may be the density of medium surrounding the nucleon,
\textit{i.e.}, the nucleon density in the nucleus without the contribution
from the nucleon under consideration. Since the calculation of this
density is qualitatively unobvious, then, to the first
approximation, we may restrict our consideration for the $A \geq 4$
nuclei simply to the highest nucleon density in the nucleus,
max$(\rho(r))$. All the nuclei, for which $S_{L,0} < 1$ have been
previously obtained, have max$(\rho(r)) > 0.16$~fm$^{-3}$
\cite{{15},{27},{28},{29}}, whereas in the $^6$Li case we have
max$(\rho(r)) =  0.15$~fm$^{-3}$ \cite{17}. From the comparison
between $S_{L,0}$ and max$(\rho(r))$ of the  nuclei under
consideration it follows that the critical density value, over which
nucleon modification takes place, is $\rho_c \approx
0.15$~fm$^{-3}$. The hypothesis of the relationship between nucleon
modification and nucleon distribution in the nucleus has been
described in detail in paper~\cite{30}.

In conclusion, we note that, as it can be seen from item {\bf{5B}},
of great importance are the experimental data on $S_L(q)$ of the
$^6$Li nucleus at $q > 1.6$~fm$^{-1}$. As should the experiment at
the Jefferson Lab~\cite{11} confirm the effect of undersaturation of
CS rules, then it would be exceptionally interesting to carry out
measurements for obtaining $S_L(q)$ of the $^6$Li nucleus at high
momentum transfers and, possibly, to perform similar measurements on
$^7$Li and $^9$Be nuclei, where the nucleon density is relatively
low. The results of these experiments would be the basis for drawing
important conclusions about nucleon modification in the atomic
nucleus.

\vspace{1cm}
% Non-BibTeX users please use

\end{multicols}
\end{document}